# Bonding characteristics of the interfacial buffer layer in epitaxial graphene via density functional theory

Alana Okullo[1], Heather M. Hill[2], Albert F. Rigosi[2], Angela R. Hight Walker[2], Francesca Tavazza[3], and Sugata Chowdhury[3,4]

[1]*Department of Physics and Astronomy, University of San Francisco, San Francisco, CA 94117, United States*

[2]*Physical Measurement Laboratory, National Institute of Standards and Technology (NIST), Gaithersburg, MD 20899, United States*

[3]*Material Measurement Laboratory, National Institute of Standards and Technology (NIST), Gaithersburg, MD 20899, United States*

[4]*Department of Physics and Astronomy, Howard University, Washington, DC 20059, United States*

ABSTRACT: Monolayer epitaxial graphene is an appropriate candidate for a wide variety of electronic and optical applications. One advantage of growing graphene on the Si face of SiC is that it develops as a single crystal, as does the layer underneath, commonly referred to as the interfacial buffer layer. The properties of this supporting layer include a band gap, making it of interest to groups seeking to build devices with on-off capabilities. In this work, using density functional theory, we have calculated the bonding characteristics of the buffer layer to the SiC substrate beneath. These calculations were used to determine a periodic length between the covalent bonds acting as anchor points in this interface. Additionally, it is evident that the formation of these anchor points depends on the lattice mismatch between the graphene layer and SiC.

# I. INTRODUCTION

Graphene has been demonstrated to exhibit desirable electrical properties [1-3]. Epitaxially grown graphene (EG) from 4H-SiC substrates continues to show promise as a method to obtain homogeneous material, which can be fabricated into devices with lateral dimensions on the millimeter and centimeter scale, a strongly applicable quality in the field of metrology [4-10]. Even with all of its exciting properties, graphene is still inherently limited when it comes to applications in the semiconductor industry requiring a band gap.

There has been a recent interest in understanding properties of the interfacial buffer layer (IBL), given that its growth and appearance are graphene-like, with the notable difference being that the two-dimensional honeycomb lattice is partially bound to the SiC substrate beneath [11-14]. This IBL has a small band gap and has been well-characterized by methods including scanning tunneling microscopy, low energy electron diffraction, and angle-resolved photoemission spectroscopy [15-20]. A variety of theoretical studies have also been performed on the electronic structure of the IBL in 6H-, 4H-, and 3C-SiC substrates [21-26].

In this work, various results of density functional theory calculations (DFT) are presented, including the generalized gradient approximation (GGA) and a revised Perdew-Burke-Ernzerhof functional (known as PBEsol) for 2H-SiC and 4H-SiC substrates. The predicted $k$-points at which periodic covalent bonding occurs, as well as the thickness of the SiC substrate at which the change in energy per atom converges are also presented. With the unit cell stretched at 8.37 %, we investigate the impacts of mechanical strain on the bonding properties of the IBL to the SiC.

## II. METHODS

### A. Sample Preparation

Several IBL samples were prepared by first performing a full EG growth on square SiC chips diced from 100 mm 4H-SiC(0001) semi-insulating wafers (CREE[see notes]). The exact parameters and procedures for the EG growth have been well documented in related work [12, 27-32]. To summarize the growth steps, heating and cooling rates were approximately 1.5 °C/s and the following steps were taken: (1) Substrate was cleaned at 1080 °C in a forming gas environment (96 % Ar, 4 % $H_2$ by volume) at 100 kPa for 30 minutes, (2) chamber was evacuated and flushed with 100 kPa Ar from a 99.999 % liquid Ar source, and (3) chamber was heated to 1900 °C.

### B. Density Functional Theory

Both 2H-SiC and 4H-SiC substrates were investigated using DFT to determine the lattice parameter, SiC layer spacings up to the IBL, and the relevant bond lengths. Calculations were carried out in plane-wave self-consistent field (PWSCF) code [33-35], using the projector augmented wave (PAW) method within the GGA [36]. For the electronic structure calculations, an 8x8x8 $k$-point grid and a 13x13x1 $k$-point grid was used for the unit cell and slab calculations, respectively. To investigate the general optical properties, a 5x5x1 $k$-point grid was used, and the Liouville-Lanczos approach was applied to linear-response time-dependent DFT, similarly applied to PWSCF, to calculate the dielectric function of the IBL [37-38]. Furthermore, the extracted band gap from these calculations yields approximately 0.3 eV, which is comparable to the result found in the recent IBL work [13].

### III. CONVERGENCE AND STRUCTURAL RESULTS

Several convergence tests were performed to determine the thickness of SiC at which an optimized model could appear, allowing one to use the same k-point grid parameters for further DFT calculations of the IBL. Convergence tests were performed using 4H-SiC as the substrate. Each layer within that substrate was 1.008 nm thick, and two layers were considered in this test. With a 2.016 nm thickness, the energy change per atom in SiC converges for the system for each test. This result indicates that it is an acceptable value for the substrate thickness while one evaluates the interactions between the IBL and substrate. At a k-point value of 13x13x1, the energy of the system converges to an asymptotic value. The final selected parameters will define the model used to explore the bonding mechanism between the IBL and the SiC. Though one could select even higher thicknesses and k-point grids for incrementally better models, the computational demands do not justify such selections.

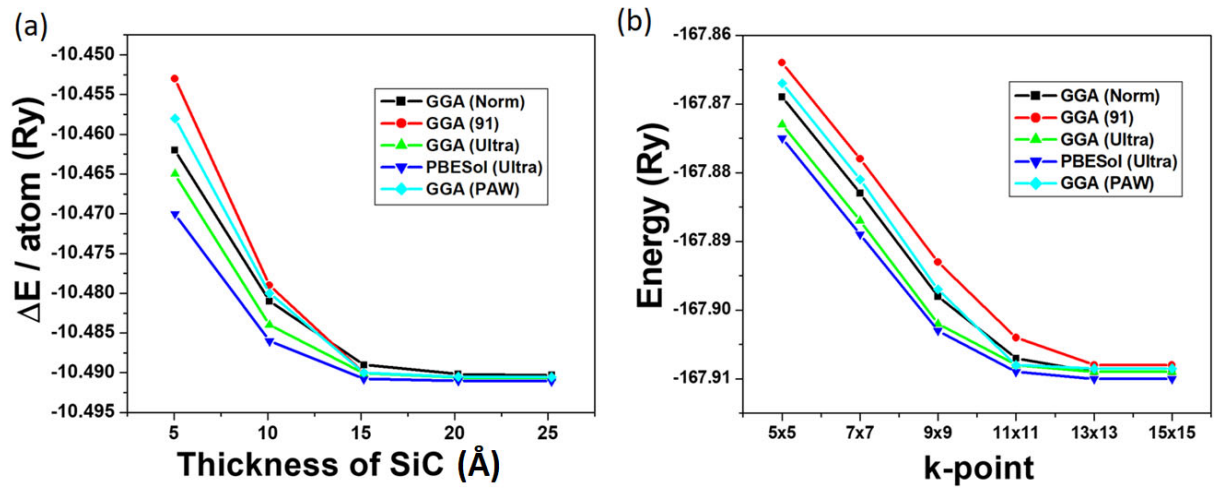

FIG. 1. Two major convergence tests are shown to exemplify how the optimal model is selected for describing the IBL/SiC interface. Within each test, several calculation methods are implemented in the hope that an agreement between all of them would indicate a likelihood that the IBL/SiC system is being accurately described. (a) The substrate thickness converged and prompted the selection of approximately 2 nm for the optimal system. (b) The optimal k-point grid selected based on these tests was 13x13x1. Here, 1 Ry = 13.605698 eV.

| Method | c (Å) (Thickness) | Si-C (Å) (Bond) | a Å(Lattice) |
|---|---|---|---|
| Starting | 20.16 | 1.868 | 3.076 |
| GGA (Norm) | 20.241 | 1.892 | 3.077 |
| LDA (Norm) | 20.215 | 1.975 | 3.036 |
| GGA (91) | 20.17 | 1.907 | 3.101 |
| LDA (PW) | 20.064 | 1.958 | 3.028 |
| PBESol (Ultra) | 20.148 | 1.945 | 3.081 |
| PBESol (PAW) | 20.163 | 1.907 | 3.088 |
| GGA (Ultra) | 20.159 | 1.901 | 3.102 |
| LDA (Ultra) | 20.092 | 1.937 | 3.013 |
| GGA (PAW) | 20.123 | 1.891 | 3.069 |
| LDA (PAW) | 20.104 | 1.928 | 3.013 |
| OptB88 | 20.085 | 1.919 | 3.035 |

FIG. 2. The 4 types of GGA calculations from the convergence tests yielded several results for 4H-SiC, including the thickness, bond length, and lattice parameter.

| Method | c (Å) (Thickness) | Si-C (Å) (Bond) | a Å(Lattice) |
|---|---|---|---|
| Exp (This Work) | 18.25 | 1.868 | 3.076 |
| GGA (Norm) | 18.226 | 1.875 | 3.077 |
| LDA (Norm) | 17.939 | 1.712 | 3.036 |
| GGA (91) | 18.146 | 1.817 | 3.101 |
| LDA (PW) | 17.882 | 1.652 | 3.028 |
| PBESol (Ultra) | 18.017 | 1.783 | 3.081 |
| PBESol (PAW) | 18.013 | 1.729 | 3.088 |
| GGA (Ultra) | 18.125 | 1.826 | 3.102 |
| LDA (Ultra) | 18.012 | 1.791 | 3.013 |
| GGA (PAW) | 18.21 | 1.848 | 3.068 |
| LDA (PAW) | 18.13 | 1.819 | 3.032 |
| OptB88 | 18.106 | 1.803 | 3.095 |

FIG. 3. The 4 types of GGA calculations from the convergence tests yielded several results for 2H-SiC, including the thickness, bond length, and lattice parameter.

As determined earlier, the 13x13x1 k-point grid was used for additional calculations involving the interactions between the IBL and SiC. These calculations were carried out in a similar fashion as before (PWSCF code, PAW, and GGA). Due to the limited accuracy of the generalized gradient approximation (GGA-DFT) to describe van der Waals forces, a newly developed van der Waals density functional was applied with C09 exchange (*vdW-DF-C09*) [39]. And within the GGA, ultrasoft pseudopotentials were implemented [40], as well as norm-conserving pseudopotentials [41]. The kinetic energy cutoff of plane-wave expansion was taken to be 544.23 eV (40 Ry). All of the geometric structures are fully relaxed until the force on each atom is less than 0.001 eV/nm, and the energy-convergence criterion is $1x10^{-6}$ eV.

Geometric parameters such as the lattice constant, bond length, and intralayer distance obtained for bulk and the (0001) surface of 4H-SiC (IBL) are summarized in Table 2-SM and are in good agreement with the experiment in this work and previous work [42, 43]. Graphene-absorbed 4H-SiC(0001) surfaces display a wide large variety of different reconstructions such as $\sqrt{3}x\sqrt{3}$ R30, 4x4, and $6\sqrt{3}x6\sqrt{3}$ R30 (with respect to the SiC 1x1 surface cell) [18, 21]. The latter two commensurate structures, 4x4 and $6\sqrt{3}x6\sqrt{3}$ R30, are too large for reasonable calculations. Therefore, in this work we have used $\sqrt{3}x\sqrt{3}$ R30 reconstruction, which corresponds to a 2x2 IBL cell. Applied strain on the IBL is more than 8 % to make it commensurate with 4H-SiC (0001) surface. A sufficiently large vacuum (20 Å) in the vertical direction is used to avoid the interaction between neighboring supercells. An example supercell is shown in Fig. 4.

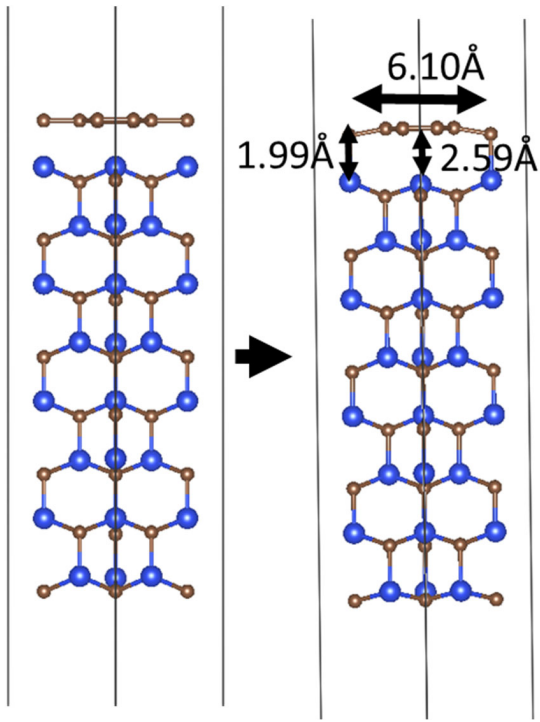

FIG. 4. Supercell of the IBL and SiC interface. DFT calculations within the ab initio supercell plane-wave approach are utilized [44]. The supercell's IBL C atoms (beige) bond to the Si atoms (blue) below and are constructed to be consistent with recent experimental observations [13].

|  | a | c | C-C in Graphene | $d_{\text{G-SiC}}$ |
|---|---|---|---|---|
| **Experimental** | 5.33 | 20.16 | - | 1.89/2.80 |
| **DFT** | 5.29 | 20.15 | 1.55 | 1.99/2.59 |

FIG. 5. Optimized geometric structure parameters of the IBL on the Si-face 4H-SiC (0001). $d_{\text{G-SiC}}$ is the distance from the IBL and the Si-face of 4H-SiC (0001). Experimental data is from Ref. [45]. All the values are given in angstroms (Å).⊥

The structure that represents the IBL is shown in Fig. 6. Optimizing the structure parameters yields a system similar to the one reported in Ref. 13. Furthermore, the electronic structure calculated from this system yields a band gap of 0.3 eV, similar to the value obtained in Ref. 13.

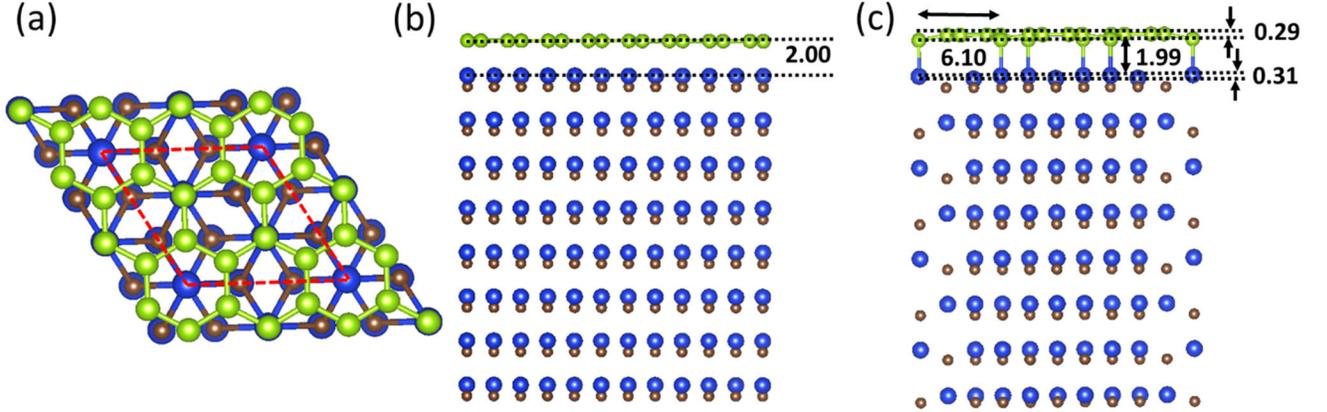

FIG. 6. (a) The top view of the IBL structure on the Si-face. Green, gray, and blue spheres represent IBL C atoms, SiC C atoms, and Si atoms, respectively. The unit cell is highlighted by red dashed lines. (b) The crystal structure of the IBL on 4H-SiC is shown before optimization. (c) The crystal structure of the IBL on 4H-SiC is shown after optimization. All values are given in angstroms (Å). $\perp$

As shown in Figure 6 (b), before the structure is optimized, the IBL layer is seemingly parallel to the SiC substrate. After structural optimization, as shown in Fig. 6 (c), a periodic formation of covalent bonding appears (6.1 Å) and a decrease in bond length from 2.00 Å to 1.99 Å.

## IV. CONCLUSIONS

The optical and electrical properties of the graphene-like interfacial buffer layer have significant promise in the semiconductor industry. Previously, the dielectric function had been reported by our group. In this work, we report the bond length of the system before and after optimization along with DFT results. The wave-like bonding characteristics observed gives insight into this graphitic system and is applicable to devices based on SiC.

## ACKNOWLEDGMENTS AND NOTES

Financial support from the REU program at Howard University (NSF Awards PHY-1659224 and PHY-1950379) is gratefully acknowledged by A.O. The authors thank Dr. Prabhakar Misra for directing the REU program.

Commercial equipment, instruments, and materials are identified in this paper in order to specify the experimental procedure adequately. Such identification is not intended to imply recommendation or endorsement by the National Institute of Standards and Technology or the United States government, nor is it intended to imply that the materials or equipment identified are necessarily the best available for the purpose. The authors declare no competing financial interest.

$^\perp$ *Though not an SI unit, this unit is permitted for expressing data by the International Committee for Weights and Measures (CIPM) and the National Institute of Standards and Technology.*

REFERENCES

[1] A.K. Geim and K.S. Novoselov, Nat. Mater. **6**, 183 (2007).

[2] K.S. Novoselov, A.K. Geim, S.V. Morozov, D. Jiang, Y. Zhang, S.V. Dubonos, I.V. Grigorieva, and A.A. Firsov, Science **306**, 666 (2004).

[3] K.S. Novoselov, V.I. Fal'ko, L. Colombo, P.R. Gellert, M.G. Schwab, and K. A. Kim, Nature **490**, 192 (2012).

[4] T.J.B.M. Janssen, A. Tzalenchuk, R. Yakimova, S. Kubatkin, S. Lara-Avila, S. Kopylov, and V.I. Fal'ko, Phys Rev B **83**, 233402 (2011).

[5] A.F. Rigosi, N.R. Glavin, C.-I. Liu Y. Yang, J. Obrzut, H.M. Hill, J. Hu, H.-Y. Lee, A.R. Hight Walker, C.A. Richter, R.E. Elmquist, and D.B. Newell, Small **13**, 1700452 (2017).

[6] A. Tzalenchuk, S. Lara-Avila, A. Kalaboukhov, S. Paolillo, M. Syväjärvi, R. Yakimova, O. Kazakova, T.J.B.M. Janssen, V. Fal'ko, and S. Kubatkin, Nat Nanotechnol. **5**, 186 (2010).

[7] A. F. Rigosi, R. E. Elmquist, Semicond. Sci. Technol. **34**, 093004 (2019).

[8] R. Ribeiro-Palau, F. Lafont, J. Brun-Picard, D. Kazazis, A. Michon, F. Cheynis, O. Couturaud, C. Consejo, B. Jouault, W. Poirier, and F. Schopfer, Nature Nanotechnol. **10**, 965 (2015).

[9] J. Hu, A. F. Rigosi, M. Kruskopf, Y. Yang, B.-Y. Wu, J. Tian, A. R. Panna, H.-Y. Lee, S. U. Payagala, G. R. Jones, M.E. Kraft, D.G. Jarrett, K. Watanabe, T. Taniguchi, R. E. Elmquist, and D. B. Newell, Sci. Rep. **8**, 15018 (2018).

[10] T. Oe, A. F. Rigosi, M. Kruskopf, B. Y. Wu, H. Y. Lee, Y. Yang, R. E. Elmquist, N. H. Kaneko, D. G. Jarrett, IEEE Trans. Instrum. Meas. **69**, 3103-8 (2019).

[11] M. S. Nevius, M. Conrad, F. Wang, A. Celis, M. N. Nair, A. Taleb-Ibrahimi, A. Tejeda, and E. H. Conrad, Phys. Rev. Lett. **115**, 136802 (2015).

[12] H. M. Hill, A. F. Rigosi, S. Chowdhury, Y. Yang, N. V. Nguyen, F. Tavazza, R. E. Elmquist, D. B. Newell, A. R. Hight Walker. Phys. Rev. B **96**, 195437 (2017).

[13] M.N. Nair, I. Palacio, A. Celis, A. Zobelli, A. Gloter, S. Kubsky, J.-P. Turmaud, M. Conrad, C. Berger, W. de Heer, E.H. Conrad, A. Taleb-Ibrahimi, and A. Tejada, Nano Lett. **17**, 2681 (2017).

[14] A. F. Rigosi, H. M. Hill, N. R. Glavin, S. J. Pookpanratana, Y. Yang, A. G. Boosalis, J. Hu, A. Rice, A. A. Allerman, N. V. Nguyen, C. A. Hacker, R.E. Elmquist, A. R. Hight Walker, D. B. Newell. 2D Mater. **5**, 011011 (2017).

[15] G.M. Rutter, N.P. Guisinger, J.N. Crain, E.A.A. Jarvis, M.D. Stiles, T. Li, P.N. First, and J.A. Stroscio, Phys. Rev. B **76**, 235416 (2007).

[16] P. Mallet, F. Varchon, C. Naud, L. Magaud, C. Berger, and J.-Y. Veuillen, Phys. Rev. B **76**, 041403 (2007).

[17] C. Riedl, U. Starke, J. Bernhardt, M. Franke, and K. Heinz, Phys. Rev. B **76**, 245406 (2007).

[18] I. Forbeaux, J. M. Themlin, and J. M. Debever, Phys. Rev. B **58**, 16396 (1998).

[19] T. Ohta, A. Bostwick, J. L. McChesney, T. Seyller, K. Horn, and E. Rotenberg, Phys. Rev. Lett. **98**, 206802 (2007).

[20] S. Y. Zhou, G.-H. Gweon, J. Graf, A. V. Fedorov, C. D. Spataru, R. D. Diehl, Y. Kopelevich, D.-H. Lee, S. G. Louie, and A. Lanzara, Nat. Phys. **2**, 595 (2006).

[21] F. Varchon, R. Feng, J. Hass, X. Li, B. N. Nguyen, C. Naud, P. Mallet, J.-Y. Veuillen, C. Berger, E. H. Conrad, and L. Magaud, Phys. Rev. Lett. **99**, 126805 (2007).

[22] E. Lampin, C. Priester, C. Krzeminski, and L. Magaud, J. Appl. Phys. **107**, 103514 (2010).

[23] X. Peng and R. Ahuja, Nano Lett. **8**, 4464 (2008).

[24] M. Inoue, H. Kageshima, Y. Kangawa, and K. Kakimoto, Phys. Rev. B **86**, 085417 (2012).

[25] A. Mattausch and O. Pankratov, Phys. Status Solidi B **245**, 1425 (2008).

[26] S. Kim, J. Ihm, H.J. Choi, and Y.-W. Son, Phys. Rev. Lett. **100**, 176802 (2008).

[27] M. Kruskopf, A. F. Rigosi, A. R. Panna, D. K. Patel, H. Jin, M. Marzano, M. Berilla, D. B. Newell, and R. E. Elmquist, IEEE Trans. Electron Dev. **66**, 3973 (2019).

[28] T. Seyller, J. Phys. Condens. Matter **16**, S1755 (2004).

[29] M. Kruskopf, A. F. Rigosi, A. R. Panna, M. Marzano, D. Patel, H. Jin, D. B. Newell, and R. E. Elmquist, Metrologia **56**, 065002 (2019).

[30] M. Kruskopf, D. M. Pakdehi, K. Pierz, S. Wundrack, R. Stosch, T. Dziomba., M. Götz, J. Baringhaus, J. Aprojanz, and C. Tegenkamp, 2D Mater. **3**, 041002 (2016).

[31] A. F. Rigosi, C.-I Liu, B.-Y. Wu, H.-Y. Lee, M. Kruskopf, Y. Yang, H. M. Hill, J. Hu, E. G. Bittle, J. Obrzut, A. R. Hight Walker, R. E. Elmquist, and D. B. Newell. Microelectron. Eng. **194**, 51 (2018).

[32] A. F. Rigosi, D.K. Patel, M. Marzano, M. Kruskopf, H. M. Hill, H. Jin, J. Hu, A.R. Hight Walker, M. Ortolano, L. Callegaro, C.-T. Liang, and D. B. Newell, Carbon **154**, 230 (2019).

[33] P. Hohenberg and W. Kohn, Phys. Rev. **136**, B864 (1964).

[34] W. Kohn and L. J. Sham, Phys. Rev. **140**, A1133 (1965).

[35] P. Giannozzi, S. Baroni, N. Bonini, M. Calandra, R. Car, C. Cavazzoni, D. Ceresoli, G. L. Chiarotti, M. Cococcioni, and I. Dabo, J. Phys.: Condens. Matter **21**, 395502 (2009).

[36] J. P. Perdew, K. Burke, and M. Ernzerhof, Phys. Rev. Lett. **77**, 3865 (1996).

[37] I. Timrov, N. Vast, R. Gebauer, and S. Baroni, Comput. Phys. Commun. **196**, 460 (2015).

[38] I. Timrov, N. Vast, R. Gebauer, and S. Baroni, Phys. Rev. B **88**, 064301 (2013).

[39] V. R. Cooper, Phys. Rev. B **81**, 161104 (2010).

[40] D. Vanderbilt, Phys. Rev. B **41**, 7892 (1990).

[41] H. M. Hill, S. Chowdhury, J. R. Simpson, A. F. Rigosi, D. B. Newell, H. Berger, F. Tavazza, A. R. Hight Walker, Phys. Rev. B **99**, 174110 (2019).

[42] M. E. Levinshtein, S. L. Rumyantsev, and M. S. Shur, *Properties of Advanced Semiconductor Materials: GaN, AlN, InN, BN, SiC, SiGe* (John Wiley & Sons, Hoboken, NJ, 2001).

[43] H. Schulz and K. Thiemann, Solid State Commun. **32**, 783 (1979).

[44] M. C. Payne, M. P. Teter, D. C. Allan, T. Arias, and J. Joannopoulos, Reviews of modern physics **64**, 1045 (1992).

[45] M. Stockmeier, R. Müller, S. Sakwe, P. Wellmann, and A. Magerl, Journal of Applied Physics **105**, 033511 (2009).